\documentclass[aps,twocolumn,showpacs,preprintnumbers,nofootinbib,prl,amsfonts,amssymb,amsmath,superscriptaddress]{revtex4-2}

\makeatletter
\def\l@subsubsection#1#2{}
\def\l@subsubsubsection#1#2{}
\makeatother

\usepackage{comment}
\usepackage{graphicx,amsthm,epsfig,epsf,fixmath}
\usepackage[usenames]{color}
\usepackage[dvipsnames]{xcolor}
\usepackage{epstopdf}

\usepackage{aas_macros}
\usepackage{bm}
\usepackage{dcolumn}
\usepackage{lipsum}
\usepackage{latexsym}
\usepackage{rotating}
\usepackage{longtable}

\usepackage{enumerate}
\usepackage{tensor,multirow}
\usepackage{url}
\usepackage[linktocpage]{hyperref}
\usepackage[caption=false]{subfig}
\usepackage[USenglish]{babel}
\usepackage{microtype}

\newcommand{\dd}{\mathrm{d}}

\newcommand{\llangle}{\langle\langle}
\newcommand{\rrangle}{\rangle\rangle}

\newcommand{\new}[1]{{{#1}}} %

\defcitealias{Cannizzaro:2020uap}{Paper~I}

\renewcommand{\sec}[1]{\textit{#1.---}}

\begin{document}

\title{Relativistic perturbation theory for black-hole boson clouds}

\author{Enrico Cannizzaro}
\email{enrico.cannizzaro@uniroma1.it}
\affiliation{Dipartimento di Fisica, ``Sapienza'' Universit\`a di Roma \& Sezione INFN Roma1, Piazzale Aldo Moro
5, 00185, Roma, Italy}
\author{Laura Sberna}
\email{laura.sberna@aei.mpg.de}
\affiliation{Max Planck Institute for Gravitational Physics (Albert Einstein Institute) Am Mu\"{u}hlenberg 1, 14476 Potsdam, Germany}
\author{Stephen R. Green}
\email{stephen.green2@nottingham.ac.uk}
\affiliation{School of Mathematical Sciences, University of Nottingham\\ University Park, Nottingham NG7 2RD, United Kingdom}
\author{Stefan Hollands}
\email{stefan.hollands@uni-leipzig.de}
\affiliation{Institut f\"ur Theoretische Physik, Universit\"at Leipzig,
  Br\"uderstrasse 16, D-04103 Leipzig, Germany}
\affiliation{Max Planck Institute for Mathematics in the Sciences, Inselstrasse 16,
  D-04109 Leipzig, Germany}

\begin{abstract} 
    We develop a relativistic perturbation theory for scalar clouds around rotating black holes. We first introduce a relativistic product and corresponding orthogonality relation between modes, extending a recent result for gravitational perturbations. We then derive the analog of time-dependent perturbation theory in quantum mechanics, and apply it to calculate self-gravitational frequency shifts. This approach supersedes the non-relativistic ``gravitational atom'' approximation, brings close agreement with numerical relativity, and has practical applications for gravitational-wave astronomy.
\end{abstract}

\maketitle

\sec{Introduction}Fundamental bosonic fields are ubiquitous in extensions of General Relativity and the Standard Model. In black hole (BH) spacetimes, perturbations by massless bosonic fields are well known to be described by a series of damped sinusoids called quasi-normal modes (QNMs)~\cite{Berti:2009kk, Kokkotas_1999}. Unlike normal modes, which exist for conservative systems and have purely real spectrum, QNMs appear in dissipative systems and have complex frequencies $\omega=\omega_\text{R}+i\omega_\text{I}$, with the imaginary part setting their decay time. For BHs, dissipation arises due to radiation of the field through the horizon and away to infinity.

Massive fields around BHs admit an additional class of solutions known as quasi-bound states (QBSs). Whereas QNMs are radiative solutions, with frequency $|\omega|>\mu$, where $\mu$ is the field mass, QBSs are spatially confined by the Yukawa suppression and have $|\omega|<\mu$. %
Thus, QBSs do not radiate at infinity, although they still dissipate through the horizon. For spinning BHs, these modes can also undergo superradiant amplification, leading to the well-known superradiant instability (see, e.g.,~\cite{Brito:2015oca}). For astrophysical BHs, this process is efficient for \mbox{$\mu \approx 10^{-20}$--$10^{-10}~\text{eV}$,} leading to the formation of a macroscopic boson cloud and the spin-down of the BH. This phenomenon translates into potentially observable signatures, such as gaps in the BH spin-mass (Regge) plane, gravitational wave emission from the condensate \new{(when the bosonic field is real)}, or signatures in binary systems \cite{Baryakhtar:2017ngi, Cardoso:2020hca, Hannuksela:2018izj, Brito:2017zvb, Brito:2017wnc, Tsukada:2020lgt, Brito:2020lup, Pani:2012vp, Baumann:2018vus, Baumann:2019ztm, Tomaselli:2023ysb, Cole:2022fir, Siemonsen:2019ebd, Isi:2018pzk, Baumann:2022pkl}. Superradiant instabilities, therefore, represent a powerful probe of ultralight bosons beyond the Standard Model, such as axions or dark photons.

Given these (and other) prospects for deviations from linear mode evolution, there is considerable interest in calculating nonlinear perturbative effects involving QBSs or QNMs~\cite{Cardoso:2020hca,Siemonsen:2022yyf,Cheung:2022rbm,Mitman:2022qdl,Brito:2023pyl}. However, due to the non-Hermiticity of the system, the spectral theorem does not guarantee the orthogonality or completeness of these modes---which moreover often diverge at the BH horizon or infinity---so it is not clear \emph{a priori} how to incorporate them into a perturbative framework.

For QBSs, the problem can be simplified using the ``gravitational atom'' or ``hydrogenic'' approximation. Indeed, at leading order in the gravitational coupling $\alpha=\mu M$, where $M$ is the BH mass, and beyond the field's Compton length, $r \gg \mu^{-1}$, QBSs reduce to eigenfunctions of the hydrogen atom Hamiltonian. In this limit, the ingoing condition at the BH horizon is replaced by a regularity condition at the origin~\cite{Detweiler:1980uk,Rosa:2009ei,Pani:2012bp,Baumann:2018vus}. Thus, a ``hydrogenic'' inner product $( \  \cdot \ , \
\cdot \ )_{\rm hyd}$ can be defined, in analogy to quantum mechanics, and mode orthogonality is guaranteed by the spectral theorem in the absence of dissipative boundaries.\footnote{The same is not true for QNMs, which still radiate to infinity.}

The hydrogenic approximation (and its relativistic corrections~\cite{Baumann:2018vus, Berti:2019wnn}) has been widely used to compute various perturbative corrections to the linear problem~\cite{Baumann:2018vus,Baryakhtar:2020gao,Baumann:2021fkf,Baumann:2022pkl,Tong:2022bbl,Siemonsen:2022yyf}. For instance, to leading order, a potential term $\delta V$ arising from, e.g., a binary companion, or a quartic self-interaction, gives rise to level mixing through the matrix element $(\Psi_{n \ell m}, \delta V \Psi_{n' \ell' m'} )_{\rm hyd} $~\cite{Baumann:2018vus}. The self-gravity of the state also gives rise to a shift in the mode frequency, proportional to the matrix element $(\Psi_{n \ell m}, \delta V \Psi_{n \ell m} )_{\rm hyd}$~\cite{Baryakhtar:2017ngi,Baryakhtar:2020gao}. However, this approximation has two drawbacks: it breaks down for higher values of $\alpha$, and it does not take into account the dissipative nature of the problem. To accurately model the phenomenology of massive fields around black holes, we require a \emph{relativistic} perturbative framework, based on an appropriate notion of orthogonality between the modes.

In this \emph{Letter,} we introduce a bilinear form for massive scalar fields in Kerr to take the place of the hydrogenic inner product in fully relativistic calculations. Under this bilinear form, which is a natural extension of the gravitational bilinear form of Ref.~\cite{Green:2022htq}, Kerr QNMs and QBSs are truly orthogonal---for all values of $\alpha$. The product reduces to the hydrogenic inner product in the limit $\alpha\rightarrow0$, but it is also applicable in the relativistic regime, and forms the basis for a relativistic perturbation theory in terms of modes.

Using the relativistic product, we derive the analog of time-dependent perturbation theory in quantum mechanics for the scalar field. As an application, we calculate the leading relativistic frequency shift due to the self-gravity of a superradiant mode, and we find a significant improvement over the hydrogenic approximation when comparing to previously-published numerical-relativity results~\cite{Tsukada:2020lgt}, improving the agreement \new{by a factor of $2.5$} even at $\alpha=0.4$. Our product therefore opens a new path to accurate nonlinear mode calculations.

We use $G_{\rm N}=c=\hbar=1$ units throughout.

\sec{Bilinear form for massive scalars}We first extend the bilinear form of \cite{Green:2022htq} to scalar \emph{massive} perturbations of Kerr and prove the orthogonality of scalar modes with both quasinormal and quasibound asymptotic conditions. 

The Kerr line element for a black hole of mass $M$ and spin parameter $a$ is given by
\begin{align}\label{eq:BL Kerr met}
{\rm d}s^2 = &-\left(1-\frac{2Mr}{\Sigma}\right) \dd t^2 - \frac{4Mar\sin^2\theta}{\Sigma} \dd t\dd\phi \nonumber \\&+ \frac{\Sigma}{\Delta}\dd r^2 + \Sigma \dd\theta^2 + \frac{\Lambda}{\Sigma} \sin^2\theta \dd\phi^2,
\end{align} 
in Boyer-Lindquist coordinates, where $\Delta=r^2+a^2-2Mr$,  $\Sigma=r^2+a^2\cos^2\theta$, $\Lambda = (r^2 + a^2)^2 - \Delta a^2 \sin^2\theta$.
We denote the event horizon (the greater root $r_\pm$ of $\Delta$) by $r_+$ and define the tortoise coordinate, ${\rm d}r/{\rm d}r_*=\Delta/(r^2+a^2)$.

\new{The Lagrangian density for the complex Klein-Gordon equation on a Kerr background (which coincides with the Teukolsky equation $\mathcal{O}\Phi = 0$ for a spin $s=0$ complex massive field~\cite{Teukolsky:1973ha}) reads
\begin{equation}\label{eq:KGmass}
   L = -\sqrt{-g}(g^{ab} \nabla_a \Phi^* \nabla_b \Phi + \mu^2 \Phi^* \Phi),
\end{equation}
}
where $\mu$ is the mass. A product between two solutions of the Klein-Gordon equation can be built as follows. Start from a ``base'' product (related to the symplectic form),
\begin{equation}
    \Pi_\Sigma[\Phi_1,\Phi_2] = \int\limits_{\Sigma}  (\Phi_1 \nabla_a \Phi_2-\Phi_2 \nabla_a \Phi_1) n^a {\rm d}V,
\end{equation}
where $\Sigma$ is a time slice with unit normal $n^a$. One can easily verify that, if $\Phi_1,\Phi_2$ are solutions, the product is conserved (i.e., independent of $\Sigma$) and that it is $\mathbb{C}$-linear in both entries, or bilinear. 

Reference~\cite{Green:2022htq} showed that one can build, from this base product, an infinite number of conserved quantities by inserting symmetry operators of the equation of motion. In Kerr, one can make use of the symmetry operators associated with the time-translation and $\phi$ rotation isometries, $\mathcal{L}_t$ and $\mathcal{L}_\phi$, as well as with the Killing tensor of the spacetime. One can also use the symmetry operator associated with the $t$--$\phi$ spacetime symmetry, $\mathcal{J}$, whose action on a scalar field simply takes $t\rightarrow-t$ and $\phi\rightarrow-\phi$. Note that the Teukolsky operator and the $t$--$\phi$ reflection operator commute on $s=0$ Weyl scalars, $\mathcal{O}\mathcal{J} = \mathcal{J} \mathcal{O}$.

The product relevant for the orthogonality relation can be built from the $t$--$\phi$ reflection operator~\cite{Green:2022htq}. For scalar massive (or massless) perturbations \emph{with compact support} it is given by
\begin{equation}\label{eq:bilinear_form_tphi}
\langle\langle \Phi_1,\Phi_2 \rangle\rangle = \Pi_\Sigma[\mathcal{J}\Phi_1,\Phi_2] \, .
\end{equation}
In Boyer-Lindquist coordinates, the bilinear form reads 
\begin{align}\label{eq:bilinearform_J}
   \langle\langle \Phi_1,\Phi_2 \rangle\rangle 
   =\int\limits_{r_+}^{\infty} {\rm d}r \int\limits_{S^2}^{} {\rm d}^2\Omega  &\bigg[\frac{2 M r a}{\Delta} \left(\mathcal{J}\Phi_{1}\partial_\phi\Phi_2 -\Phi_2\partial_\phi\mathcal{J}\Phi_{1} \right) \nonumber \\
    &+\frac{\Sigma }{\Delta} \left( r^2 + a^2 + \frac{2 M r a^2}{\Sigma}\sin^2 \theta\right) \nonumber\\
    & \times \left( \mathcal{J}\Phi_{1}\partial_t\Phi_2 -\Phi_2\partial_t\mathcal{J}\Phi_{1} \right)
    \bigg] ,
\end{align}
where ${\rm d}^2\Omega = \sin \theta \, {\rm d}\theta {\rm d}\phi$.
In addition to being bilinear and conserved, one can easily prove, in analogy to Ref.~\cite{Green:2022htq}, that
\begin{enumerate}
\item the bilinear form is symmetric, $\langle\langle \Phi_1,\Phi_2 \rangle\rangle = \langle\langle \Phi_2,\Phi_1 \rangle\rangle$; and
\item the time-translation symmetry operator is symmetric with respect to the bilinear form, $\langle\langle {\mathcal L}_{t}\Phi_1,\Phi_2 \rangle\rangle = \langle\langle \Phi_1, {\mathcal L}_{t}\Phi_2 \rangle\rangle$.\label{item:Lt}
\end{enumerate}

\sec{Extension to mode solutions}Quasinormal and quasibound states are mode solutions of the Teukolsky equation, $\Phi_{\ell m\omega} = e^{-i\omega t+i m\phi} R_{\ell m\omega}(r) S_{\ell m\omega}(\theta)$, where $S_{\ell m\omega}$ are the $s=0$ spin-weighted spheroidal harmonics with angular numbers $\ell$, $m$~\cite{Teukolsky:1973ha} and the radial solution can be defined in terms of an asymptotic series involving a three-term recursion relation~\cite{Leaver:1985ax,Konoplya:2006br}. 
The modes are required to be regular at the horizon, $\Phi \sim e^{-ik_Hr_*}$ as $r_*\to -\infty$, where $k_H =\omega-m\Omega_H$ and $\Omega_H$ is the angular frequency of the outer horizon
$\Omega_H=a/(2Mr_+)$. At infinity, the two families satisfy
\begin{align}
    \Phi &\sim  r^{-1} e^{ik r_*}, \qquad  r_*\to\infty \quad {\rm (QNMs),}\\
    \Phi  &\sim  r^{-1} e^{-ik r_*}, \qquad  r_*\to\infty \quad {\rm (QBSs),}
\end{align}
where $k = \sqrt{\omega^2-\mu^2}$.

\begin{figure*}[ht]
  \centering
  \includegraphics[trim={12.5cm 5.3cm 11.cm 4.4cm},clip,width=0.34\textwidth]{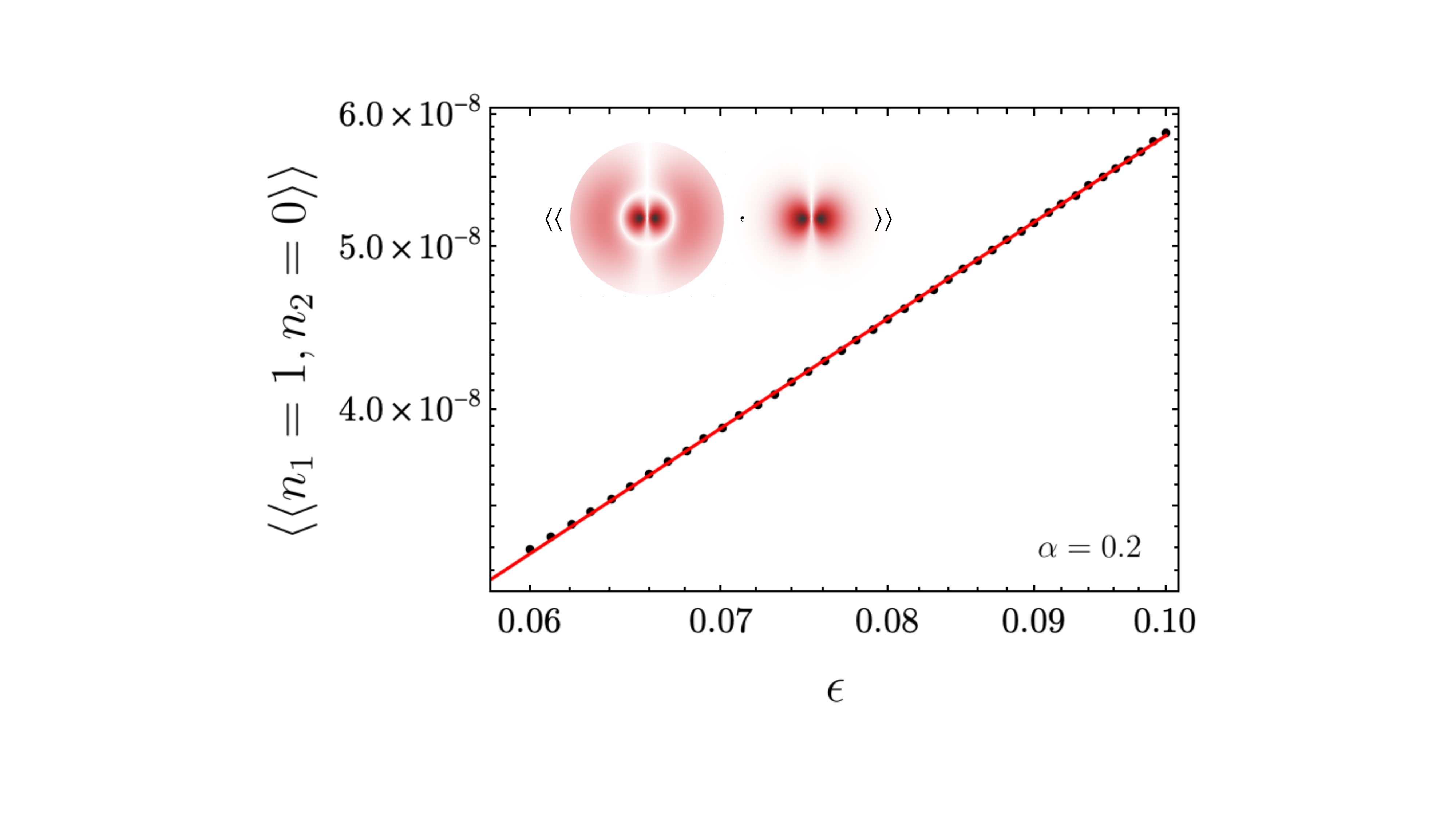}
  \includegraphics[trim={1.2cm 0.38cm 0.cm 0.cm},clip,width=0.322\textwidth]{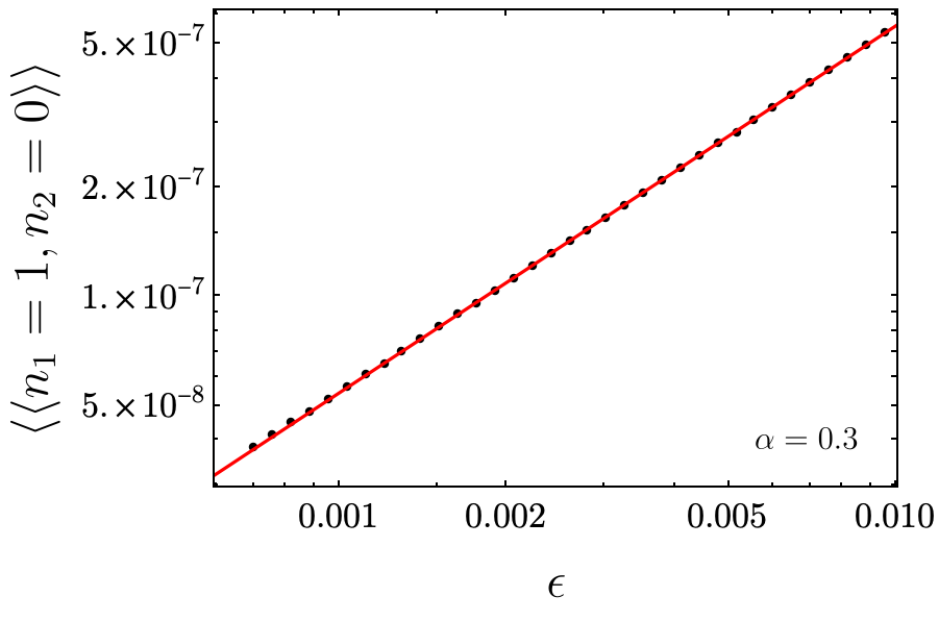}
  \includegraphics[trim={1.2cm 0.38cm 0.cm 0.cm},clip,width=0.32\textwidth]{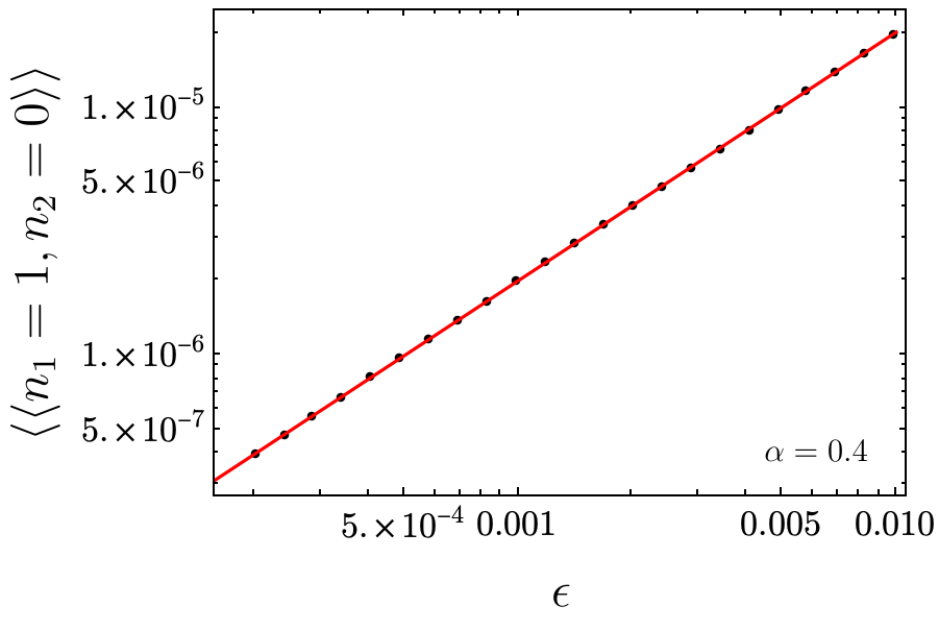}
  \caption{The relativistic product between two $\ell=m=1$ QBSs in Schwarzschild, as a function of the counter-term regularization point $\epsilon=\bar{r}/r_+-1$, for different scalar field masses. The red curve is a power-law fit, showing convergence to zero. In the top-left corner, we show the absolute value of the radial mode-functions around the BH. Modes are normalized to have $\llangle n,n \rrangle=1$ in the regularization limit.}
  \label{fig:orthogonality}
\end{figure*}

Because the radial solutions have non-compact support, and for $\omega_\text{I}<0$ actually diverge as $r_*\to -\infty$ (QNMs and QBSs) and as $r_*\to +\infty$ (QNMs), we must find a suitable, finite extension of the bilinear form \eqref{eq:bilinear_form_tphi}. 
In analogy with Ref.~\cite{Green:2022htq}, we extend the definition of the bilinear form to a complex radial integration contour $\mathcal{C}$, such that the radial integral is absolutely convergent. We define the bilinear form over a pair of QNMs or QBSs with complex frequencies $\omega_1$, $\omega_2$ by integrating over a complex $r_*$ contour such that
\begin{equation}\label{eq:contour_qbs}
\arg r_* + \arg (\omega_1+\omega_2) = - \pi/2, \quad r_* \to -\infty \, ,
\end{equation}
and running along the real axis elsewhere. If the product is over one or two QNMs, we also take
\begin{equation}\label{eq:contour_qnms}
\arg r_* + \arg (\pm k_1\pm k_2) =\pi/2, \quad r_* \to \infty\, ,
\end{equation}
where the plus (minus) sign holds for QNMs (QBSs).

Explicitly, the bilinear form on modes reads
\begin{equation}
   \langle\langle \Phi_1,\Phi_2 \rangle\rangle_{\text{modes}} = i\delta_{m_1m_2} e^{-i(\omega_1-\omega_2)t} \int\limits_\mathcal{C}^{} {\rm d}r \frac{K}{\Delta}  R_1 R_2, 
\end{equation}
where
\begin{align}
   K(r)={}&
    \alpha_{12} (r^2+a^2)^2(\omega_2+\omega_1)-2 M r a \alpha_{12} (m_1+m_2) \nonumber\\
    &- \gamma_{12} (\omega_2+\omega_1)a^2 \Delta(r),
    \\
    \alpha_{12} ={}& 2 \pi \int\limits_0^\pi {\rm d}\theta \, \sin \theta \, S_1(\theta) S_2(\theta), \\
    \gamma_{12} ={}& 2 \pi \int\limits_0^\pi {\rm d}\theta \, \sin^3 \theta \, S_1(\theta) S_2(\theta).
\end{align}
Note that, as demonstrated for Kerr QNMs in Ref.~\cite{Green:2022htq}, this product can be used to project initial data onto modes, resulting in the known mode excitation coefficients~\cite{Leaver:1986gd,Nollert:1998ys,Berti:2006wq}. In the hydrogenic limit, this reduces to the familiar inner product on the (real) hydrogenic mode functions,
\begin{align}
   \langle\langle \Phi_1,\Phi_2 \rangle\rangle &\to \delta_{m_1m_2} \int\limits_{0}^{\infty} {\rm d}r \, r^2 \, R_1 R_2(r) \int\limits_0^\pi {\rm d}\theta \, \sin{\theta} \,
 S_1 S_2(\theta) \nonumber\\
 &\equiv ( \Phi_1 , \Phi_2 )_{\rm hyd},
\end{align}
up to an overall factor. In this limit, no regularization is required.

For QBSs in Schwarzschild, it is convenient to adopt an alternative regularization involving counter-term subtraction~\cite{Sberna:2021eui}. This is particularly useful when mode solutions are only known numerically \new{and thus cannot be easily continued into the complex $r_*$-plane.} For Schwarzschild, the integrals over $r$ and $\theta$ factorize, and the latter gives rise to the usual orthogonality relation for spherical harmonics. The radial integration can then be regularized by subtracting suitable counter-terms,
\begin{align}\label{eq:qbs-product}
  & \llangle \Phi_1 , \Phi_2\rrangle_{\text{Schwarzschild QBS}}\nonumber\\
  &= i \delta_{m_1m_2} \delta_{l_1l_2}\left(\omega_1+\omega_2 \right)
   \lim_{\bar{r}_{*}\rightarrow - \infty} \left[\int_{\bar{r}_*}^{\infty} dr_* X_{1}( r_*')X_{2}( r_*')\right.\nonumber\\
   &\qquad\left.+\frac{i}{\omega_{1}+\omega_{2}} X_{1}(\bar{r}_*)X_{2}(\bar{r}_*) + \mathcal{O}(r_*^{-1}) \right],
\end{align}
where $X(r)=rR(r)$. For long-lived states ($M|\omega_I| \ll 1$), only the leading counter-term is needed to make the product finite. Further details of the counter-term subtraction method, including a discussion of higher-order counter-terms, are provided in the Supplemental Material. Note that this method works for QBSs, since regularizing the QNM divergence at infinity would require an infinite series of subtractions.%

\sec{Mode orthogonality}With the finite bilinear form in hand, from property \ref{item:Lt} we obtain 
\begin{equation}
    (\omega_1-\omega_2) \llangle \Phi_1 , \Phi_2 \rrangle=0
\end{equation}
for a pair of QNMs or QBSs with frequencies $\omega_1$, $\omega_2$. Then, either $\llangle \Phi_1 , \Phi_2 \rrangle = 0$ or $\omega_1 = \omega_2$, proving that QNMs and QBSs are orthogonal. In particular, modes of the two families are also mutually orthogonal.

We now numerically compute the product \eqref{eq:qbs-product} between two QBSs in Schwarzschild with different radial numbers\footnote{The product between states with different $\ell,m$ numbers is trivial, as their orthogonality in Schwarzschild follows from the orthogonality of the spherical harmonics. The orthogonality between different $\ell$ modes becomes non trivial in Kerr~\cite{Green:2022htq}, due to the spin-weighted spheroidal harmonics.} $n$. We do so in  the hydrogenic ($\alpha=M\mu \ll 1$) and relativistic ($\alpha\simeq 1$) regimes. To compute the QB frequencies and radial solutions, we use the Leaver continued fraction method~\cite{Leaver:1985ax}. We perform product integrals \eqref{eq:qbs-product} numerically using %
\textsc{Mathematica}. %

Figure~\ref{fig:orthogonality} shows the product between the $\ell=m=1$ fundamental mode and the first overtone as a function of the integral regulator $\epsilon=\bar{r}/r_+-1$. 
Different panels span the hydrogenic regime and the relativistic regime. The product between the two modes goes to zero as a power-law as $\epsilon \rightarrow 0$ in all cases, confirming numerically the orthogonality to a precision of order $10^{-7}$. For higher values of $\alpha$, we are able to probe the integral for smaller $r$ due to better convergence resulting from milder divergences at the horizon. We obtain similar results also for higher radial overtones.

\sec{Relativistic perturbation theory}\new{We now describe our relativistic approach to compute transitions between modes due to a perturbation. 
To emphasize the similarity to ordinary Schr\" odinger perturbation theory in quantum mechanics, we work 
in the Hamiltonian formulation, writing the metric in Arnowitt-Deser-Misner form $g^{ab} = -N^{-2}(t^a-N^a)(t^b-N^b)+h^{ab}$ (see App.~E of 
\cite{Wald:1984rg}), assumed to be some perturbation of Kerr. Starting from the Lagrangian \eqref{eq:KGmass}
we introduce the momentum $\Pi = N^{-1}\sqrt{h}(t^a - N^a)\nabla_a\Phi$ and the Hamiltonian, leading to 
equations in first order form,
\begin{equation}
   \left(
   \begin{matrix}
    \dot \Phi\\
    \dot \Pi
    \end{matrix}
    \right) \equiv
{\mathcal L}_t  \left(
   \begin{matrix}
     \Phi\\
    \Pi
    \end{matrix}
    \right) = H  \left(
   \begin{matrix}
     \Phi\\
    \Pi
    \end{matrix}
    \right) ,
\end{equation}
where 
\begin{equation}\label{eq:perturbed_Teukolski}
H   \equiv
      \left(
    \begin{matrix}
    N^a D_a & N/\sqrt{h}\\
    \sqrt{h}(D^a N D_a -N\mu^2) & D_a N^a
    \end{matrix}
    \right).
\end{equation}
In phase-space notation, the relativistic product takes the form $\llangle (\Phi_1,\Pi_1)^{\rm T}, (\Phi_2,\Pi_2)^{\rm T} \rrangle = \int_{\mathcal{C}} (\Phi_1\circ {\mathcal J} \Pi_2+\Pi_1 \circ {\mathcal J} \Phi_2)\,{\rm d}^3 x$.

For a general perturbation%
, $H$ is time-dependent. We
make an ansatz for the column vector $F=(\Phi,\Pi)^{\rm T}$ associated with a solution in terms of a superposition of modes with time-dependent amplitudes,\footnote{
Considering the $\Pi$ and $\Phi$ components separately, 
one sees that consistency of \eqref{eq:linearcomb} 
requires $|\dot c_q/c_q| \ll |\omega_q|$ and $a/M \lesssim 1$. 
}
\begin{equation}
\label{eq:linearcomb}
    F(t)=\sum_q c_q(t) F_{0q}(t),
\end{equation}
where $F_{0q}$ are the QB and QN modes of the unperturbed problem with Hamiltonian $H_0$, i.e.,~$H_0 F_{0q} \equiv {\mathcal L}_t F_{0q} = -i\omega_q F_{0q}$, 
so that $F_{0q}(t)$ has harmonic $e^{-i\omega_q t}$ time dependence.

We decompose the Hamiltonian as $H=H_0+\delta H$, where the subscript 0 denotes quantities associated with the Klein-Gordon equation in the Kerr metric. 
The scheme rests on the facts that $\llangle F_{0q},F_{0q'} \rrangle = \delta_{qq'}$ and that $H_0$ is symmetric\footnote{The symmetry $\llangle F_1, H_0 F_2 \rrangle = \llangle H_0 F_1, F_2 \rrangle$ can be shown by integration by parts, and in contrast to property \ref{item:Lt} does not require $F_i$ to be solutions.} relative to our relativistic product on two-vector states $F_i$. A standard calculation mirroring quantum mechanics then gives the perturbation series for the 
time-dependent excitation coefficients,
\begin{equation}\label{eq:pertseries}
    \dot{c}_n \llangle \Phi_n, \Phi_n\rrangle = \sum_q c_q \llangle F_{0n},\delta H(t) \, F_{0q} \rrangle.
\end{equation}

If $\delta H$ is approximately $t$-independent, we have an (approximate) perturbed QN or QB mode $F = F_{0} + \delta F$, defined by
the ``eigenvalue equation'' $H F = -i(\omega_0 + \delta \omega)F$ and appropriate boundary conditions, with a frequency shift $\delta \omega$. Taking an inner product $\llangle F_{0}, \ \cdot \ \rrangle$ with the unperturbed QNM or QBS and going through exactly the same steps as in ordinary time-independent quantum mechanics
perturbation theory immediately yields the usual formula,
\begin{equation}
\label{eq:frequencyshift}
-i\delta \omega = \frac{\llangle F_0, \delta H F_0 \rrangle}{\llangle F_0, F_0 \rrangle},
\end{equation}
at first perturbation order. Substituting this back into the eigenvalue equation and taking an inner product $\llangle F_{0q}, \ \cdot \ \rrangle$ with all unperturbed QNM or QBS 
$F_{0q}$ orthogonal to $F_0$ then gives
\begin{equation}
\delta F_0 = \sum_q \frac{\llangle F_{0q}, \delta H F_0 \rrangle}{-i\llangle F_{0q}, F_{0q} \rrangle \, (\omega_0-\omega_{0q})} F_{0q},
\end{equation}
using \eqref{eq:pertseries} at first order. In the \emph{Supplemental Material}, we also derive the perturbation equations for the excitation coefficients in the second-order formalism.

\sec{Frequency shift due to self-gravity}We apply our relativistic perturbative framework to calculate the frequency shift $\delta \omega_n$ of an (unstable) mode $\Phi_n$ close to the superradiant 
bound in Kerr due to its self-gravity.
We assume that the squared amplitude $A^2$ of the mode and the rotation parameter $a/M$ are both relatively small and neglect effects that are not linear
in these quantities. We show in the \emph{Supplemental Material} that, under these assumptions, the perturbed metric 
$\delta g^{ab}=g^{ab}-g^{ab}_0$ can be written in the form $\delta g^{ab} \approx 
-\delta [N^{-2} (t^a - N^a)(t^b-N^b)]$ where $\delta N^a \approx N_0^a \delta N/N_0$. Following~\cite{Siemonsen:2022yyf}, we therefore take a semi-Newtonian, approximation for the gravitational potential 
sourced by a mode. With this, the perturbed Hamiltonian of the scalar field is $\delta H \approx \delta V H_0$, where $\delta V = \delta N/N_0$ is given approximately by
\begin{equation}
    \delta V(r) \approx - \mu^2 \left[ \frac{1}{r}\int\limits_{r_+}^{r} {\rm d}^3 r' \, |\Phi_n|^2  + 
    \int\limits_{r}^{\infty} {\rm d}^3 r' \, \frac{ |\Phi_n|^2}{r'} \right],
\end{equation}
where the integration is carried out over flat space and we have taken the leading order (spherical) multipole.

To estimate the correction $\delta\omega_n$ to the mode frequency $\omega_n$ in Kerr we use \eqref{eq:frequencyshift}, with $\delta H \approx \delta V H_0$ and $H_0(\Phi_n, \Pi_n)^{\rm T} = -i\omega_n(\Phi_n, \Pi_n)^{\rm T}$. Reverting to 1-component form, we get
\begin{equation}\label{eq:fre_shift_bilinear}
    \frac{\delta\omega_n}{\omega_n} \approx  \frac{\llangle \Phi_n, \delta V \Phi_n\rrangle}{\llangle \Phi_n, \Phi_n\rrangle} .
\end{equation}
}
This approach is similar in spirit to that outlined in Refs.~\cite{Zimmerman:2014aha,Mark:2014aja,Hussain:2022ins,Brito:2023pyl}. In the nonrelativistic limit, this formula reduces to that found in Refs.~\cite{Baryakhtar:2020gao,Siemonsen:2022yyf}. Note that superradiantly unstable QBSs, which have $\omega_I>0$ and decay at infinity, have no divergence at the horizon and therefore require no regularization of the product.%

\new{For completeness, we also write the equation for the time evolution of the excitation coefficients, better suited to when the perturbation $\delta H \approx \delta V H_0$ is time-dependent,
\begin{equation}
    \dot{c}_n \llangle \Phi_n, \Phi_n\rrangle = - i\sum_q \omega_q  \llangle \Phi_n, \delta V \Phi_q\rrangle.
\end{equation}
}

We now calculate numerically the frequency shift~\eqref{eq:fre_shift_bilinear} for superradiant modes with $\ell=m=1$. For a given coupling $\alpha$, we set the BH spin to be close to the superradiant bound $m \Omega_H \gtrsim \omega_R$, the same setup as~\cite{Siemonsen:2022yyf}. For this application, we use the \texttt{Black Hole Perturbation Toolkit} to compute the modes' spin-weighted spheroidal harmonics~\cite{BHPToolkit}.

\begin{figure}[t]
\centering
\includegraphics[width=\linewidth]{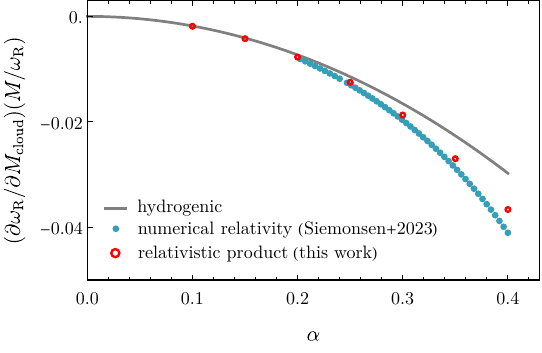}
\caption{Frequency shift due to the self-gravity of a superradiant mode in Kerr ($\ell=m=1,$ $n=0$). We compare our result based on the relativistic product with the hydrogenic approximation, and with the fully relativistic (numerical) frequency shift from Ref.~\cite{Siemonsen:2022yyf}. For the analytic results, we plot $\delta \omega/ M_{\rm cloud}$, which should be a good approximation of the derivative for small cloud masses.}
\label{fig:freq_shift}
\end{figure}

In Fig.~\ref{fig:freq_shift}, we compare for several $\alpha$ our perturbative calculation of $\delta \omega/ M_{\rm cloud}$ against the numerical relativity estimate of $\partial \omega /\partial M_{\rm cloud}$ from \cite{Siemonsen:2022yyf}. We find excellent agreement, including significant improvement over the hydrogenic approximation, which begins to fail around $\alpha \simeq 0.3$. For $\alpha = 0.4$, the error is reduced from 28\% \new{to 11\%}. %
The remaining disagreement is likely due to the approximation that $\delta \omega$ is linear in the cloud mass ($\partial \omega /\partial M_{\rm cloud} \simeq \delta \omega/ M_{\rm cloud}$), to our semi-Newtonian approximation \new{for the perturbed equation, and to the monopolar approximation of the Newtonian potential.}

In the \emph{Supplemental Material}, we include another example, calculating relativistic matrix elements of tidal perturbers~\cite{Baumann:2018vus}, and again find $O(10\%)$ corrections to the hydrogenic approximation. This example is relevant for gravitational wave signals from extreme or intermediate mass-ratio binaries (see also \cite{Brito:2023pyl}).

\sec{Conclusions}In this study, we introduced a bilinear form for massive scalar-field perturbations of Kerr and showed that modes are orthogonal with respect to this product. Our bilinear form replaces the standard quantum mechanics inner product---often employed in a hydrogenic approximation---making no assumption on the strength of the gravitational coupling $\alpha$. We also introduced an approach to compute perturbative corrections to mode evolution due to a \new{perturbation}, and applied this to recover frequency shifts due to the self-gravity of a superradiant state. For large values of $\alpha$, accurate results were previously only obtainable using numerical relativity.

Our bilinear form and perturbative framework have both conceptual and practical importance.
Other applications could be to compute corrections due to self-interaction terms such as quartic potentials \cite{Baryakhtar:2020gao}, or in the sine-Klein-Gordon equation for the QCD axion \cite{Yoshino:2012kn}. In future work, we also hope to explore transitions between QN and QB modes, and to rigorously derive angular selection rules for massive perturbations in Kerr using the bilinear form.

Another natural extension would be to generalize our product to massive spin-1 fields. This scenario presents a number of difficulties as the Proca equation is not separable using the standard Teukolsky formalism. Nevertheless, an ansatz yielding separability of the Proca equation in Kerr spacetime was recently discovered \cite{Lunin:2017drx, Frolov:2018ezx}, and could allow for a generalization of the bilinear form. 

Finally, in the context of BH binaries, the gravitational product~\cite{Green:2022htq} could be used with the second-order Teukolsky equation~\cite{Green:2019nam,Spiers:2023cip} to estimate nonlinear corrections to the BH ringdown. This could be used to inform waveform development and address recent questions on nonlinear effects during the ringdown~\cite{Giesler:2019uxc,Baibhav:2023clw,Nee:2023osy,Cheung:2022rbm,Mitman:2022qdl}. We hope to report on these interesting problems in the future.

\begin{acknowledgments}
\sec{Acknowledgments}The authors would like to thank T.~May, N.~Siemonsen and W.~East for sharing the numerical relativity data for Fig.~\ref{fig:freq_shift}. The authors would also like to thank S.~Dolan for suggesting to explore the orthogonality of quasibound states at the 2022 Capra Meeting, and R.~Brito, V.~Cardoso, F.~Duque, T.~Spieksma, G.~M.~Tomaselli and S.~V\"{o}lkel for their helpful comments during the preparation of the manuscript. This work makes use of the Black Hole Perturbation Toolkit.  
\end{acknowledgments}

\bibliography{Ref}

\clearpage
\begin{center}
  \large
  \textbf{Supplemental Material}
\end{center}

\section{Counter-term subtraction method}

\label{sec:regular}In certain applications, e.g., when mode solutions are only known numerically, the complex-contour regularization can be difficult to implement in practice. 
Here we introduce an alternative regularization based on the subtraction of divergent boundary terms at the horizon \cite{Sberna:2021eui}. %

Consider for simplicity the bilinear form (4) in Schwarzschild. It is immediate to see that in this case the integrals in $r$ and $\theta$ factorize, with the latter reducing to the orthogonality condition for spherical harmonics,  
\begin{align}
\label{eq:bilinearformnospin}
   \langle\langle \Phi_1,\Phi_2 \rangle\rangle = i\delta_{m_1m_2} \delta_{l_1l_2}\left(\omega_1+\omega_2 \right) %
   \int\limits_{r_+}^{\infty} {\rm d}r \, 
    \frac{r^2}{f} R_1 R_2,
\end{align}
where $f =1 - 2 M/r$.
It is convenient to define the scalar function $R(r)=X(r)/r$, so that the radial integral becomes simply
\begin{equation}
\langle\langle \Phi_1,\Phi_2 \rangle\rangle \sim \int {\rm d}r \, f^{-1} X_1 X_2 = \int {\rm d}r_* \, X_1 X_2.
\end{equation}
In the near-horizon limit, the solution $X$ can be described by the series expansion 
\begin{equation}
\label{eq:horizonexpantion}
    X(r)\sim \sum_{n=0}^{+\infty} X^n_{r_+} \sim \sum_{n=0}^{+\infty} e^{-i \omega r_*}(r-r_+)^n b_n(\omega).
\end{equation}
At leading order, the mode behaves as
\begin{equation}
    X^0_{r_+}(r) \sim e^{-i \omega r_*} \sim (r-r_+)^{r_+ \omega_I}.
\end{equation}
When the mode is stable ($\omega_I<0$) this term diverges at the horizon. To regularize this divergence, we simply subtract the near-horizon integral of the leading order term, 
\begin{align}
\label{eq:overlapintegral}
   \llangle \Phi_1 , \Phi_2\rrangle \sim
   \lim_{\bar{r}_{*}\rightarrow - \infty} \bigg[&\int\limits_{\bar{r}_*}^{\infty} {\rm d}r_* X_{1}( r_*')X_{2}( r_*')\nonumber\\
   &+\frac{i}{\omega_{1}+\omega_{2}} X_{1}(\bar{r}_*)X_{2}(\bar{r}_*) \bigg] .
\end{align}
Note that this method is suitable to regularize the horizon divergence of both (massless or massive) QNMs and QBSs, as their leading-order behavior in the near-horizon expansion coincides.

This method can be extended to regularize the divergence of any order $N=n_1+n_2$ in the horizon expansion \eqref{eq:horizonexpantion}, 
\begin{equation}
\label{eq:NLOhorizon}
    X_{r_+}^{n_1} X_{r_+}^{n_2}\sim (r-r_+)^{N} (r-r_+)^{r_+ \left( \omega_{I1}+\omega_{I2}\right)}.
\end{equation}
The $N$-th term in the expansion is regular if $r_+ \omega_{I1}+r_+ \omega_{I2}+N\ge 0$. 
However, QBSs with interesting (i.e., potentially detectable) phenomenology are long lived modes, $M|\omega_I|\ll 1$. In this limit, all terms beyond the leading order are regular, and the subtraction of the leading order divergence \eqref{eq:overlapintegral} is sufficient. %
 
The complex contour and the counter-term subtraction method give equivalent definitions of the mode product. To see this, we separate the integral into
\begin{equation}\label{eq:complex_to_boundary}
\langle\langle \Phi_1,\Phi_2 \rangle\rangle \sim \int\limits_\mathcal{C}^{} {\rm d}r_* X_1 X_2 = \int\limits_\mathcal{C_{+}} {\rm d}r_* X_1 X_2 + \int\limits_{\bar{r}_*}^{+\infty} {\rm d}r_* X_1 X_2.
\end{equation}
where $C_{+}: r_* = \bar{r}_* + \rho e^{i\beta}$ and $\beta$ is an angle in the complex plane chosen to satisfy condition \eqref{eq:contour_qbs}. Assuming the modes are very bound $M|\omega_I|\ll 1$ and in the limit of $\bar{r}_*\to -\infty$, we find that the first integral is equal to
\begin{align}
 & e^{-i (\omega_1+\omega_2)\bar{r}_*} e^{i \beta} \int\limits_{\infty}^{0} {\rm d}\rho \ e^{-i (\omega_1+\omega_2)\rho e^{i\beta}} \left[1+O(\rho^{-1})\right]\nonumber\\
 &= \frac{i}{\omega_1+\omega_2} e^{-i (\omega_1+\omega_2)\bar{r}_*}\left[1+O(\rho^{-1})\right]\nonumber\\
 &=  +\frac{i}{\omega_{1}+\omega_{2}} X_{1}(\bar{r}_*)X_{2}(\bar{r}_*) \left[1+O(\bar{r}_*^{-1})\right].
\end{align}
This is precisely the counter term defined in \eqref{eq:overlapintegral}.

\section{Covariant perturbation theory}

Instead of the Hamiltonian approach to perturbation theory outlined 
in the \emph{Letter}, we may also consider a ``covariant'' (one component) form.
As in the previous section, we consider a perturbation $g^{ab} = g^{ab}_0+\delta g^{ab}$, with correspondingly perturbed KG operator $\mathcal{O} = \mathcal{O}_0
+\delta \mathcal{O}$. As in the main text, the ansats for the solution to 
$\mathcal{O} \Phi = 0$ is approximated by a sum of QN or QB modes $\Phi_{0q}$ 
on the unperturbed spacetime, with 
$t$-dependent coefficients,
\begin{equation}
\label{eq:modesum}
    \Phi = \sum_q c_q(t) \Phi_{0q}
\end{equation}
where the dependence of $\Phi, \Phi_{0q}$ on $t$ is implicitly understood.  
Consider the quantity $\Pi_{\Sigma(t)}(\Phi_1 \circ \mathcal{J}, \Phi_2) = \llangle 
\Phi_1, \Phi_2 \rrangle_{\Sigma(t)}$ on the perturbed spacetime. Then we have 
\begin{equation}
\label{eq:master}
   \frac{\dd}{\dd t} 
   \llangle 
\Phi_1, \Phi_2 \rrangle_{\Sigma(t)}
   = \int\limits_{\Sigma(t)} [(\mathcal{O} \Phi_1 \circ \mathcal{J})\Phi_2 
   - \Phi_1 \circ \mathcal{J} (\mathcal O \Phi_2)] \, t^a \dd \Sigma_a ,
\end{equation}
which in the absence of a perturbation states that the bilinear form does not depend on the chosen Cauchy surface $\Sigma(t)$ if $\Phi_1, \Phi_2$ solve the 
unperturbed KG equation. Now we take $\Phi_1 = \Phi_{0q}$, a QN or QB mode of the unperturbed spacetime, and $\Phi_2 = \Phi$. Furthermore, instead of $\Sigma(t)$ we take one of the complex contours $\mathcal{C}(t)$ required to define the inner product for QN or QB modes, as described in the \emph{Letter}. Assuming that $|\dot c_q/c_q|\ll |\omega_q|$, we then have 
\begin{equation}
      \llangle 
\Phi_{0q}, \Phi \rrangle_{\mathcal{C}(t)} \approx c_q(t) \llangle 
\Phi_{0q}, \Phi_{0q} 
\rrangle_{\mathcal{C}(t)}.
\end{equation}
Combining with Eq. \eqref{eq:master} gives using $\mathcal{O}\Phi \approx 0$
and $\mathcal O_0 \Phi_{0q} = 0$ gives
\begin{equation}
\begin{split}
    \dot c_q(t)
   \approx& \sum_{q'} c_{q'}(t) \int\limits_{\mathcal{C}(t)} (\delta \mathcal{O} \Phi_{0q}) \circ \mathcal{J} \, \Phi_{0q'}  \, t^a \dd \Sigma_a \\
   \equiv& \sum_{q'} \delta \mathcal{O}(t)_{qq'} c_{q'}(t),
\end{split}
\end{equation}
where we defined $\delta \mathcal{O}(t)_{qq'} = \int_{\mathcal{C}(t)} (\delta \mathcal{O} \Phi_{0q}) \circ \mathcal{J} \, \Phi_{0q'}  \, t^a \dd \Sigma_a$.
This equation is solved as usual by a perturbation series similar to 
Eq. \eqref{eq:pertseries} in the \emph{Letter}
\begin{equation}
\label{eq:pertseries}
\begin{split}
    &c_q(t) = c_q(t_0) + \sum_{n>0} \ \ 
        \int\limits_{t_0<s_1<\dots<s_n<t} {\rm d}^n s \\
        \qquad 
&\delta \mathcal{O}(s_n)_{qq_n} \cdots \delta \mathcal{O}(s_2)_{q_3q_2} 
    \delta \mathcal{O}(s_1)_{q_2q_1} c_{q_1}(t_0),
    \end{split}
\end{equation}

If $\delta \mathcal O$ can be assumed to be nearly time-independent, we may obtain an expression for the frequency shift of a QN or QB mode 
$\Phi_q = \Phi_{0q} + \delta \Phi_q$ defined by 
${\mathcal L}_t \Phi_q = -i(\omega_q +\delta \omega_q) \Phi_q$ and appropriate boundary condtions. Using the properties of the bilinear form, we then have 
\begin{equation}
   \frac{\dd}{\dd t} 
   \llangle 
\Phi_{0q}, \delta \Phi_q \rrangle_{\mathcal{C}(t)}
=
   -\llangle {\mathcal L}_t
\Phi_{0q}, \delta \Phi_q \rrangle_{\mathcal{C}(t)}
   +\llangle 
\Phi_{0q}, {\mathcal L}_t \delta \Phi_q \rrangle_{\mathcal{C}(t)}.
\end{equation}
On the other hand, Eq. \eqref{eq:master} gives
\begin{equation}
   \frac{\dd}{\dd t} 
   \llangle 
\Phi_{0q}, \delta \Phi_{q} \rrangle_{\mathcal{C}(t)}
   = -\int\limits_{\mathcal{C}(t)} \Phi_{0q} \circ \mathcal{J} (\mathcal{O}_0 \delta \Phi_q) \, t^a \dd \Sigma_a .
\end{equation}
Since to first order in perturbation theory $\mathcal{O}_0 \delta \Phi_{q} + \delta \mathcal{O} \Phi_{0q}=0$, and $
\mathcal{L}_t \delta \Phi_{q} = -i(\omega_q \delta \Phi_q + \delta \omega_q \Phi_{0q})$, we easily get
\begin{equation}
-i\delta \omega_q = \frac{(\delta \mathcal{O})_{qq}}{\llangle \Phi_{0q}, \Phi_{0q}\rrangle}
\end{equation}
for the frequency shift.
In Schwarzschild, we 
can re-write $(\delta \mathcal O)_{qq} = (2\omega_q)^{-1} \llangle \Phi_{q0}, (N^2\delta \mathcal O) \Phi_{q0} \rrangle$.

\section{Perturbed scalar field Hamiltonian}

Here we estimate the perturbation $\delta H$ to the scalar field Hamiltonian due to the self-gravity of 
a mode $\Phi$ of relatively small amplitude. We work on a Kerr background with relatively small $a/M$ and 
count orders for simplicity in a dimensionless parameter $\delta \lesssim 1$ so that $a/M = O(\delta) = |\Phi|^2$, though we could 
in principle easily take $|\Phi|^2 \ll a/M$ as well. The mode is assumed to be of superradiant (unstable)
type with frequency $\omega=\omega_R+i\omega_I$ close to the superradiant bound, with
$\omega_R \lesssim m\Omega_H = O(a/M)$, field mass $\mu^2 \approx \omega_R^2$ and $M|\omega_I| \ll 1$, so the field amplitude's modulus is growing only slowly. 
This growth is neglected. An unstable mode ($\omega_I>0$) is regular at $r=r_+$ and supported far\footnote{For $\alpha = 0.1$, the peak is at $r/r_+ \approx 100$;
for $\alpha = 0.4$, the peak is at $r/r_+ \approx 7$.} ($1 \lesssim r/r_+$) from the ergo-surface.

To include the self-gravity of such a mode, we must in principle consider the full Einstein-scalar field system. This 
is difficult, so we resort to a semi-Newtonian approach. 
It is convenient to discuss this in the Hamiltonian formulation of the Einstein-complex-scalar equations, see e.g.~App.~E of \cite{Wald:1984rg} for 
details and notation. For the coupled Einstein-scalar system, we supplement the Lagrangian 
to 
\begin{equation}
L=\sqrt{-g}\left( \frac{1}{8\pi G_{\rm N}} R-g^{ab}\nabla_a \Phi^* \nabla_b \Phi-\mu^2\Phi^*\Phi\right).
\end{equation} 
To simplify the formulas and differently from the \emph{Letter}, our units are such that $8\pi G_{\rm N}=1$ in the following.
The ADM-ansatz is $g^{ab} = -N^{-2}(t^a-N^a)(t^b-N^b)+h^{ab} \equiv -n^an^b+h^{ab}$, and the ADM momentum is 
$p^{ab}=\sqrt{h}(K^{ab}-h^{ab}K)$. We write $N = N_0+\delta N$, $p^{ab} = p_0^{ab}+\delta p^{ab}$, $h_{ab}=h_{0ab}+\delta h_{ab}$ etc.~where a ``$0$'' 
refers to the Kerr quantities, and when referring to the size of such tensors, 
we mean a Cartesian coordinate system made from the Boyer-Lindquist coordinates $(r,\theta,\phi)$. 

Our aim is to solve the ADM evolution- and constraint equations approximately by a semi-Newtonian, time-independent ansatz. 
In this ansatz, we 
attempt to put $\delta h_{ab} = 0$, while keeping 
the other variables perturbatively small in $\delta$,
in such a way as to be consistent with the ADM evolution- and constraint equations. We will argue that it is possible to achieve
\begin{equation}
\begin{split}
&\delta p^{ab} = O(\delta^2), \quad \delta h_{ab} = 0, \\
&\delta N^a = O(\delta), \quad \delta N = O(\delta).
\end{split}
\end{equation}
In fact, as we shall see momentarily, consistency with the ADM evolution- and constraint equations requires that we make specific choices of $\delta N=(\delta V)N_0$, $\delta N^a=(\delta V)N^a_0 + \delta X^a$, $\delta p^{ab}$ as, respectively
\begin{equation}
\label{eq:defn}
\begin{split}
D^a \left( N_0^2 D_a \delta V \right) = N_0^2 \left(\frac{\Pi^*\Pi}{h} - \frac{\mu^2 \Phi^*\Phi}{2} \right) & \\
D^a \left( \frac{1}{N_0} D_a \delta X_b \right)  = -2{\rm Re}\Pi^* D_b \Phi/\sqrt{h} &\\
\delta p_{ab} = -\delta \left[ \frac{\sqrt{h}}{N}\left( D_{(a} N_{b)} -\frac{1}{2} h_{ab} D_c N^c \right) \right].
\end{split}
\end{equation}
The right sides of the first two equations are $O(\delta^3)$  functions varying on a dimensionless length scale 
of $1/\delta$ by our assumptions about $\Phi, \omega, \mu$. Both equations are of Laplace type (divergence form), so solutions 
are easily shown to exist and are expected to be of order $\delta V = O(\delta) = \delta N^a$ up 
to and including $r=r_+$, with derivatives falling off correspondingly faster. Thereby, 
the third equation yields $\delta p^{ab} = O(\delta^2)$.

We now show that our ans\" atze are consistent with the ADM constraint- and evolution equations at the indicated perturbative orders in $\delta$. The ADM evolution equation for $h_{ab}$
is 
\begin{equation}
\dot h_{ab} \ (=0) \ = \frac{2N}{\sqrt{h}}(p_{ab}-\tfrac{1}{2} h_{ab}p) - 2D_{(a} N_{b)}
\end{equation}
which holds in the background since Kerr is stationary. The third equation is made precisely in such a way that it holds exactly 
also for the perturbed solution. Consider next the ADM evolution equation for $p^{ab}$. For consistency, 
we need to show $\dot p^{ab} = O(\delta^2)$. This equation is lengthier and can be found e.g. in E.2.36 of App.~E of \cite{Wald:1984rg}, to which we need to add the contributions 
\begin{equation}
\begin{split}
\dot p^{ab} =& \dots + N\sqrt{h} D^{(a} \Phi^* D^{b)} \Phi - \frac{1}{2} N\sqrt{h}(D^c \Phi^* D_c \Phi \\
& \qquad + \mu^2 \Phi^*\Phi) h^{ab} + \frac{N\Pi^*\Pi}{2\sqrt{h}}.
\end{split}
\end{equation}
Using that the equation holds in the background with $\dot p^{ab}_0=0$, one easily finds $\dot p^{ab}=O(\delta^2)$.

We must also consider the Hamiltonian and momentum constraints, which are
\begin{equation}
\begin{split}
\tfrac{1}{2} ({\mathcal R} + p^{ab}p_{ab}/h - \tfrac{1}{2}p^2/h) =& \ T_{ab}n^a n^b,\\
D_a(p^{ab}/\sqrt{h}) =& \ T_{cd}n^c h^d{}_b 
\end{split}
\end{equation}
where $T_{ab}$ is the stress tensor of $\Phi$ and ${\mathcal R}_{ab}$ is the Ricci curvature of $h_{ab}$.
Taking the divergence of the ADM evolution equation for $h_{ab}$ 
and using the second equation in \eqref{eq:defn}, we see that the momentum constraint holds up to an including terms of order $O(\delta^3)$.
Taking the trace of the ADM evolution equation for $p^{ab}$ and using the first equation in \eqref{eq:defn}, we 
see that the Hamiltonian constraint holds at the same order as $\dot p$, i.e. $O(\delta^2)$, as 
is required for consistency. 

Our mode is supported relatively far from the horizon, where $N_0 \approx 1, N_0^a \approx 0, h_{ab} \approx \delta_{ab}$. 
Under these approximations $D_a \to \partial_a$ and the source of $\delta X_a$ in Eq. \eqref{eq:defn} (the momentum current) becomes 
$\approx -2\omega {\rm Im} \Phi^* D^a \Phi$. For $\omega^2 \approx \mu^2$, this term is negligible, so we may also 
neglect $\delta X^a$, and with this, $\delta N^a \approx (\delta V) N^a_0$ at the leading approximation. The remaining Eq. \eqref{eq:defn} become 
\begin{equation}
\label{eq:defn1}
\begin{split}
\vec \nabla^2 \delta V \approx & \   \frac{\mu^2 \Phi^*\Phi}{2} \approx 4\pi G_{\rm N} T_{tt}  \\
\delta p_{ab} =& \ O(\delta^3)
\end{split}
\end{equation}
where we have reinserted Newton's constant (so $\delta p^{ab}$ is seen to be even smaller than originally anticipated). 
The Poisson equation may be solved in the usual way which in the $s$-wave approximation 
gives the expression for $\delta V$ in the \emph{Letter}, the formula for $\delta N, \delta N^a$, and that of the scalar field Hamiltonian $H$,
given by $\delta H = (\delta V)H+O(\delta^2)$
at the leading approximation. Since the perturbation itself is of order $\delta \sim |\Phi|^2$, the subleading part may be dropped in 
the computation of the frequency shift $\delta \omega = O(|\Phi|^2)$.

\section{Application to tidal potential}

As another example, we consider the tidal, Newtonian potential arising from a non-spinning binary companion to a (Schwarzschild) black hole endowed with a QB state. In the hydrogenic approximation, this was shown to give rise to transitions between modes of the QB spectrum~\cite{Baumann:2018vus,Tong:2022bbl}.

A binary companion induces a perturbation in the background metric $g^{ab}=g_0^{ab}+\delta g^{ab}$. This leads to a shift in the potential of the scalar field (at leading order) $\delta V \sim \delta g_{tt}\sim \sum_{\ell_p m_p } r^{\ell_p}Y_{\ell_p m_p}$, where the subscript distinguishes the angular numbers of the perturbation \cite{Baumann:2018vus}.\footnote{In our formalism, it is possible to consistently include relativistic corrections to the tidal potential by working out the perturbed scalar field 
Hamiltonian $\delta H$ to a better approximation than given above. However, a full description of the binary system is beyond the scope of this work.}
We neglect the time dependence of the potential due to the companion's motion. This could be introduced after the calculation of the matrix element in an adiabatic approximation~\cite{Baumann:2018vus}, or taken fully into account in the relativistic matrix element.%

We compute the level mixing due to this external potential between modes with $\ell=1$, $m=1,-1$, and $n=0,1$ (or $| 211 \rangle$ and $| 31\,{-1} \rangle$ in the hydrogenic notation\footnote{In this notation, states are labelled by the three quantum numbers $|\Tilde{n}\ell m\rangle$, where $\Tilde{n}=\ell+n+1$ is the principal quantum number.}). Note that this transition is allowed in the case of tidal quadrupolar perturbations $\ell_p=2$ by the angular selection rules due to the angular dependence of the perturbed potential \cite{Baumann:2018vus}. These selection rules survive in the relativistic limit (at least in Schwarzschild) as the bilinear form reduces to the standard inner product for spherical harmonics. We will therefore focus on the radial part of the matrix element. 

In the hydrogenic approximation, the matrix elements of $\delta V$ read
\begin{equation}
\label{eq:radialintHydro}
     ( \Psi_{\tilde{n}_1\ell_1 m_1}^{\rm hyd}, \delta V \Psi_{\tilde{n}_2\ell_2 m_2}^{\rm hyd} )_{\rm hyd} \sim \int\limits_0^\infty {\rm d}r\, r^{4} R^{\rm hyd}_{\tilde{n}_1\ell_1 m_1}R^{\rm hyd}_{\tilde{n}_2\ell_2 m_2} ,
\end{equation}
where $R^{\rm hyd}_{n\ell m}$ are the hydrogenic wavefunctions,
\begin{align}
    R^{\rm hyd}_{\tilde{n}\ell m}(r)=&\sqrt{\Big(\frac{2 \mu \alpha}{\tilde{n}}\Big)^3\frac{(\tilde{n}-\ell-1)!}{2\tilde{n} (\tilde{n}+\ell)!}}\Big(\frac{2 \alpha \mu r}{\tilde{n}}\Big)^\ell e^{-\frac{\mu \alpha r}{\tilde{n}}}\nonumber \\&\times L^{2\ell+1}_{\tilde{n}-\ell-1}\Big(\frac{2 \mu \alpha r}{\tilde{n}}\Big)\, . 
\end{align}
Notice that the hydrogenic wavefunctions are everywhere regular and are integrated from the origin, as already discussed. 

\begin{figure}[t]
\centering
\includegraphics[width=0.42\textwidth]{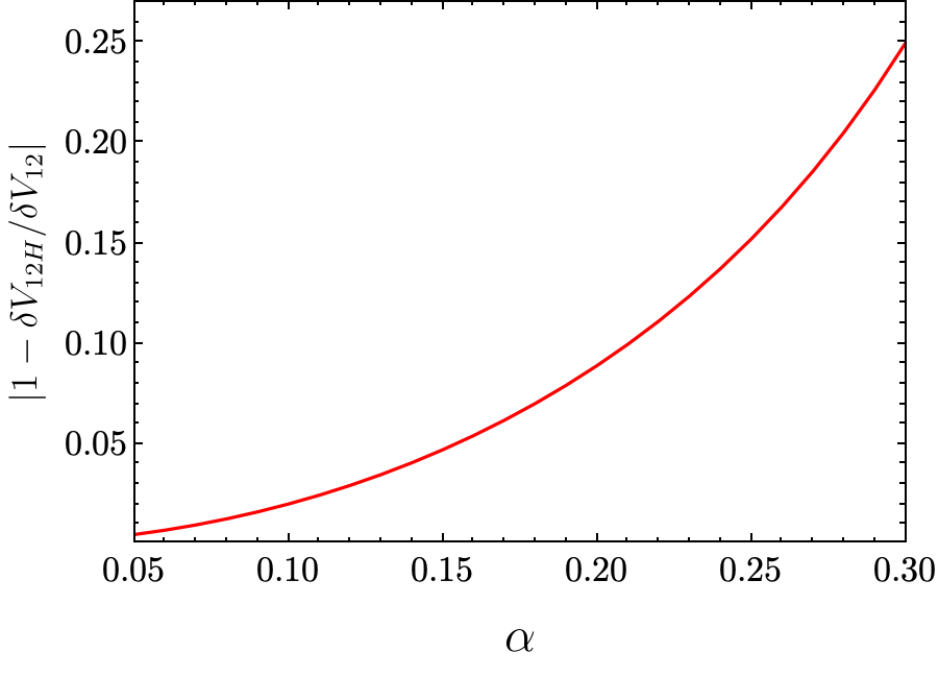}
\caption{Relative difference in the matrix element of the companion's tidal potential, computed with our bilinear form and in the hydrogenic approximation. The discrepancy increases with $\alpha$, as the hydrogenic approximation breaks down.}
\label{fig:hydroplot}
\end{figure}

The matrix element appearing in the fully relativistic perturbative expansion, on the other hand, reads
\begin{align}\label{eq:inttidal}
   \llangle \Phi_{n_1\ell_1 m_1},\delta V \Phi_{n_2\ell_2 m_2} \rrangle  \sim & \int\limits _{r_+(1+\epsilon)}^{\infty} {\rm d}r\, r^2 f(r) {X}_{1}( r'){X}_{2}( r')\nonumber \\
   &+\frac{i r_+^2}{\omega_{1}+\omega_{2}} {X}_{1}(R){X}_{2}(R)\, ,
\end{align}
Notice that the potential modifies the boundary regularization term at the horizon. In the nonrelativistic limit, the full radial solutions $X$ become real and tend to the hydrogenic approximation: $R^{\rm hyd}_{n\ell m}(r)\approx {\rm Re}(X_{n\ell m}(r))/r$. 

We evaluate the relativistic matrix element \eqref{eq:inttidal} on numerical Klein-Gordon solutions. We find that this converges to a finite nonzero value as we take the integral regulator $\epsilon\to0$, confirming that the tidal potential gives rise to level-mixing between the two modes under consideration. %

In Fig.~\ref{fig:hydroplot} we show the relative difference between the hydrogenic matrix element appearing in Ref.~\cite{Baumann:2018vus} and our fully relativistic treatment \eqref{eq:inttidal}, for a range of scalar masses $\alpha$.
While the matrix elements obtained with the two methods are overall comparable, the relative error clearly increases for higher $\alpha$, as the hydrogenic approximation starts to loose accuracy. 

Interestingly, the relative error can be as high as $10 \%$ even for values of $\alpha \simeq 0.2$, where the radial solutions are still relatively accurate (within at most $ 5\%$). %
This suggests that relativistic corrections to the mode product itself---included in our fully relativistic bilinear form, but not in the quantum-mechanics inner product---contribute significantly to the matrix element. Previous work on mode mixing focused on relativistic corrections to the decay width of the modes~\cite{Berti:2019wnn}, or resorted to fully numerical methods~\cite{Cardoso:2020hca}.

\end{document}